\documentclass[
]{ceurart}
\usepackage{listings}

\usepackage{xcolor}

\begin{document}

\copyrightyear{2025}
\copyrightclause{Copyright for this paper by its authors.
  Use permitted under Creative Commons License Attribution 4.0
  International (CC BY 4.0).}

\conference{CSEDM'25: 9th Educational Data Mining in Computer Science Education (CSEDM) Workshop,
  July 20, 2025, Palermo, Sicily, Italy}

\title{Combining Log Data and Collaborative Dialogue Features to Predict Project Quality in Middle School AI Education}

\author[1]{Conrad Borchers}[
orcid=0000-0003-3437-8979,
email=cborcher@cs.cmu.edu
]

\fnmark[1]
\address[1]{Carnegie Mellon University}

\author[2]{Xiaoyi Tian}[
orcid=0000-0002-5045-0136,
email=xtian9@ncsu.edu,]
\fnmark[1]
\address[2]{North Carolina State University}

\author[3]{Kristy Elizabeth Boyer}[%
orcid=0000-0003-3434-3450,
email=keboyer@ufl.edu,
]

\author[3]{Maya Israel}[%
orcid=0000-0003-0302-6559,
email=misrael@coe.ufl.edu,
]
\address[3]{University of Florida}

\fntext[1]{Conrad Borchers and Xiaoyi Tian contributed equally to this work.}

\begin{abstract}
Project-based learning plays a crucial role in computing education. However, its open-ended nature makes tracking project development and assessing success challenging. We investigate how dialogue and system interaction logs predict project quality during collaborative, project-based AI learning of 94 middle school students working in pairs. We used linguistic features from dialogue transcripts and behavioral features from system logs to predict three project quality outcomes: \textit{productivity} (number of training phrases), \textit{content richness} (word density), and \textit{lexical variation} (word diversity) of chatbot training phrases. We compared the predictive accuracy of each modality and a fusion of the modalities. Results indicate log data better predicts productivity, while dialogue data is more effective for content richness. Both modalities modestly predict lexical variation. Multimodal fusion improved predictions for productivity and lexical variation of training phrases but not content richness. These findings suggest that the value of multimodal fusion depends on the specific learning outcome. The study contributes to multimodal learning analytics by demonstrating the nuanced interplay between behavioral and linguistic data in assessing student learning progress in open-ended AI learning environments.

\end{abstract}


\maketitle

\section{Introduction}
 
Project-based learning (PBL) is crucial in STEM, especially in computing, where students collaboratively create artifacts \cite{pucher2011project,kokotsaki2016project}. PBL fosters computational thinking, problem-solving \cite{koh2010project,guo2020review}, and engagement \cite{koh2010project,widyaningsih2020implementation}. Effective assessment of student learning and project quality is essential for educators to provide targeted feedback \cite{wilkerson2014construction}. However, because of the iterative, open-ended nature of project development (e.g., app creation), tracking progress in PBL is challenging. Analyzing behavioral data during PBL offers insights into student learning, collaboration, and refinement strategies. Prior work has explored predicting project outcomes such as final grades \cite{oviatt2013written,yoo2014can,tian2024examining}, task performance \cite{samadi2024cultural,stewart2021multimodal}, group satisfaction \cite{acosta2024multimodal,tian2024investigating}, and engagement \cite{grafsgaard2013automatically}. In computer science education specifically, there is a growing interest in understanding more granular aspects of the project, such as the completeness of the student code traces \cite{morshed2021progression,marwan2019evaluation}, or the correctness of the steps within multi-step problem-solving episodes \cite{emerson2019predicting}. These finer-grained project quality outcomes capture learning progress during PBL, enabling educators and intelligent systems to identify struggling learners and provide timely support. Predicting project quality \textbf{proxies} \textit{during} learning processes (in addition to conventional, post-completion learning outcomes) can serve two future research purposes: First, it may inform adaptive modules that provide additional instruction based on detected areas of improvement. Second, it can inform the study of effective student collaboration by providing insights through models capturing what contributes to desirable learning process outcomes. 

Researchers have explored various data sources and analytical approaches to better understand and support student learning during PBL. A longstanding tradition in learning analytics involves inferring cognitive and metacognitive processes from log and language data \cite{borchers2024using}. While researchers have leveraged log interactions to model learning trajectories \cite{emerson2019predicting,borchers2024combining}, and language data to infer self-regulated learning strategies \cite{zhang2024using}, these modalities are rarely used for joint prediction. Integrating dialogue-based and log-based features may offer a more holistic view of student collaborative learning analytics \cite{spikol2017using,yan2024evidence}. While prior research has focused on predicting learning gains using multimodal collaboration features \cite{olsen2020temporal}, less attention has been given to the temporal dynamics of learning \cite{borchers2024using}. We address this gap by predicting the quality of individual learning events in collaborative learning, that is, the quality of training phrases while students program a chatbot. Specifically, we compare the predictive capabilities of dialogue and log data in assessing the quality of student projects in an AI learning context where students collaboratively develop chatbots. We examine whether combining different modalities of data enhances the predictive accuracy of project quality compared to using them independently:

\textbf{RQ1:} How well can student project quality be predicted from single modalities (dialogue, log data)?\\

\textbf{RQ2:} To what extent does the multimodal fusion of these data sources enhance predictive accuracy?\\

\section{Related Work}

\subsection{Multimodal Fusion and Prediction}

Multimodal approaches to mining data and modeling are increasingly common in educational data mining (EDM) \cite{sharma2020multimodal,karumbaiah2023spatiotemporal}. A recent review by Chango et al. \cite{chango2022review} traces emerging trends in the use of multimodal in typical EDM contexts, including online learning environments, classroom studies, and blended learning contexts. Data collected in these sites can range from text data to log data, physiological learner data such as eye-tracking, and spatiotemporal data such as teacher positions \cite{karumbaiah2023spatiotemporal}.

Natural language data, especially, is increasingly used for foundational EDM prediction tasks. For example, Zhang et al. \cite{zhang2024using} utilized large language model (LLM) embeddings to predict self-regulated learning stages from think-aloud protocols, illustrating the potential of text-based data in understanding cognitive processes. Scarlatos et al. \cite{scarlatos2024exploring} explored using LLM for knowledge tracing from learners' dialogue. Borchers et al. \cite{borchers2024combining} fused data from peer tutoring chat collaborations with tutoring system problem-solving logs to model learning rates in relation to different dialogue acts. However, these approaches have rarely combined language data with log data to jointly model learning.

The present study builds on this body of work by applying text mining techniques based on automated transcriptions and embedding \cite{zhang2024using} with log data for a novel prediction task: predicting learner project quality in AI education. Specifically, we incorporate transcriptions of students' collaboration as a data source in EDM, combining them with established log data to enhance predictive accuracy. By testing the combined predictive power of log data and transcription-derived features, we test whether multimodal fusion improves the prediction of learning processes during collaboration. We contribute to the ongoing discourse on the utility of multimodal data in enhancing educational predictions and insights.

\subsection{Mining Collaborative Learning Systems}

Collaborative learning systems offer opportunities to mine multimodal data from learning-system and learner-learner interactions \cite{walker2014adaptive,costley2022system}.\footnote{We define collaboration as the coordinated effort to jointly engage in problem-solving and knowledge construction \cite{roschelle1995construction}.} These systems provide insights into learning patterns through modeling tutoring rates and collaborative behaviors \cite{borchers2024combining}, which have been used for feedback tools \cite{echeverria2024teamslides,yan2024evidence}. However, selecting meaningful outcomes and learning constructs for multimodal analysis remains challenging. \textbf{The present study embraces this challenge in a novel predictive context, }where students learn to program chatbots using natural language inputs. We predict the quality of student-created training phrases during learning. Insights from this research have implications for other collaborative systems and performance prediction from collaborative characteristics \cite{borchers2024combining,olsen2020temporal,zhou2024detecting,abdelshiheed2024aligning}.

In addition to peer collaboration interactions, research has demonstrated how dialogue can be leveraged to support learner-system interactions. For instance, intelligent agents can adapt instruction based on detected learning patterns \cite{latham2010oscar}. In contrast, teachable agents provide practice problems and assessments \cite{matsuda2011learning} and build rapport with learners through lexical adaptation toward learners \cite{tian2020understanding}. 
Similarly, adaptive dialogue systems such as EER-tutor have demonstrated improved learning gains by considering prior student errors \cite{weerasinghe2010evaluating}. Finally, Gaze Tutor \cite{d2012gaze} leveraged gaze data to detect disengagement and adapt dialogue to re-engage students. The present study builds on this foundation by seeking to develop meaningful predictors of student learning performance during collaboration, making progress toward detectors that could improve adaptivity in future collaborative learning systems.

\section{Study Setup and Dataset}
\subsection{Study Context and Data Collection}

The study was conducted in a public middle school in the southeastern United States in Spring 2024. This IRB-approved study included 128 students across six class periods in a science class. Out of these, 100 consented to participate in the research, and 97 reported their demographic information: 49 identified as girls, 46 as boys, one as non-binary, and one preferred not to disclose. Further, 38 students identified as Asian, 34 as White, 20 as Black/African American, six as Hispanic/Latinx, three as Native American, five as self-described, and three preferred not to disclose. The average age was 11.7 years (SD = 0.48). 

The primary objective of this study was to engage students in core concepts of artificial intelligence (AI) and computer science (CS) through the hands-on development of conversational agents \cite{song2023ai}. The classroom study contained a 10-hour learning module covering AI fundamentals, hands-on activities, and chatbot development. The instructional content was aligned with AI Five Big Ideas in K12 \cite{touretzky2019envisioning}. Specific learning goals included (1) understanding the role of data in AI systems and determining appropriate datasets for specific AI applications; (2) describing how datasets create representations of the world for reasoning tasks; (3) training and evaluating classification and prediction models while examining their accuracy on new inputs; and (4) creating chatbots and developing natural interactions. Each learning module was designed to build students' AI literacy progressively. 

During collaborative chatbot development, students were randomly paired to create chatbots on science topics of their choice (e.g., the water cycle, climate, and living organisms). Each pair collaborated over three class sessions (40 minutes each) to develop their chatbot. We captured audio/video recordings of these collaborative sessions using recording software from their laptop. Humans transcribed the recordings through the Rev service. The resulting dialogue dataset contains 121 collaboration sessions from 47 student pairs (94 individuals). Each session contains an average of 278 utterances (SD = 108.7).

\subsection{Chatbot Development Environment}
\begin{figure*}[htp!]
    \centering
    \includegraphics[width=1\textwidth]{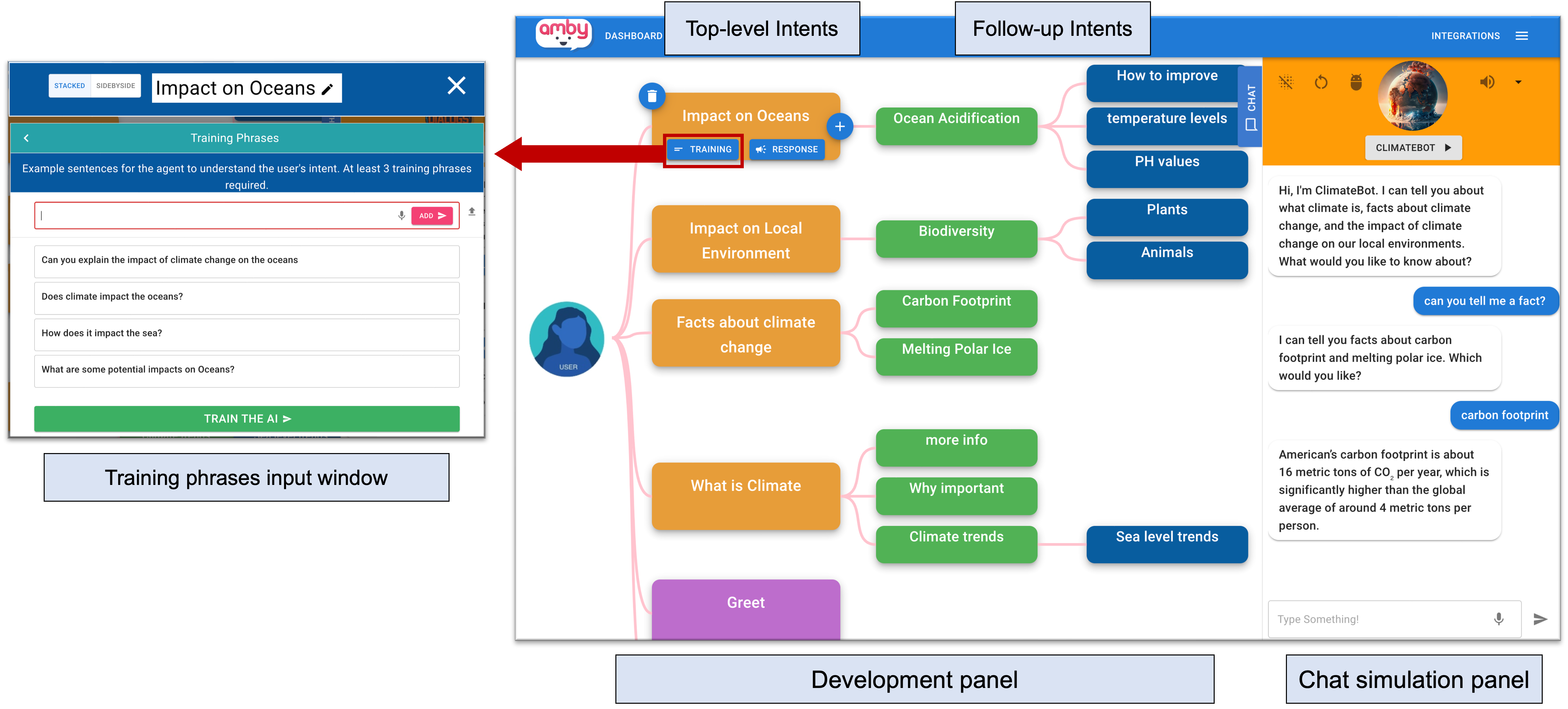}
    \caption{Chatbot development environment AMBY}
    \label{fig:amby-screenshot}
\end{figure*}

The learning environment, AMBY (Figure \ref{fig:amby-screenshot}), is a graphical interface designed for middle school students to create conversational agents and learn about AI \cite{tian2023amby}. The environment enables students to test example agents, edit agents, or create new agents. For instance, a student group might create an agent on the topic of ``climate change,'' where they define \textit{intents} (e.g., ``impact on oceans''), input \textit{training phrases} (e.g., ``Does climate change impact the oceans?''), and write corresponding \textit{responses} (e.g., ``Climate change impacts oceans through sea-level rise and ocean acidification.''). Students can test their agents by having conversations and adjusting the chatbot's voice to personalize the interaction.

Students construct and refine intents (i.e., basic chatbot development units representing the goal or purpose of a user message). In the development panel (Figure \ref{fig:amby-screenshot}, left), students can add, edit, or delete training phrases and responses. They submit training requests by clicking ``train the AI,'' which updates the chatbot's intent classification model based on the provided training data. Students worked with a partner at the same computer on iterating on the chatbot design, following the pair programming paradigm \cite{hanks2011pair}, prompting spontaneous collaboration, dialogue, and joint decision-making.

The system collects 23 types of timestamped \textbf{user interaction logs} (e.g., adding or deleting training phrases, submitting AI training requests, chat messages). Across 121 recorded collaborative sessions, students submitted an average of seven intent training requests per session, and we used the content of training phrases they submitted each time as measures of project quality (Section \ref{outcome-measure}).  

\subsection{Dataset Preparation} \label{sec:dataset-prep}

\begin{figure*}
    \centering
    \includegraphics[width=.66666\textwidth]{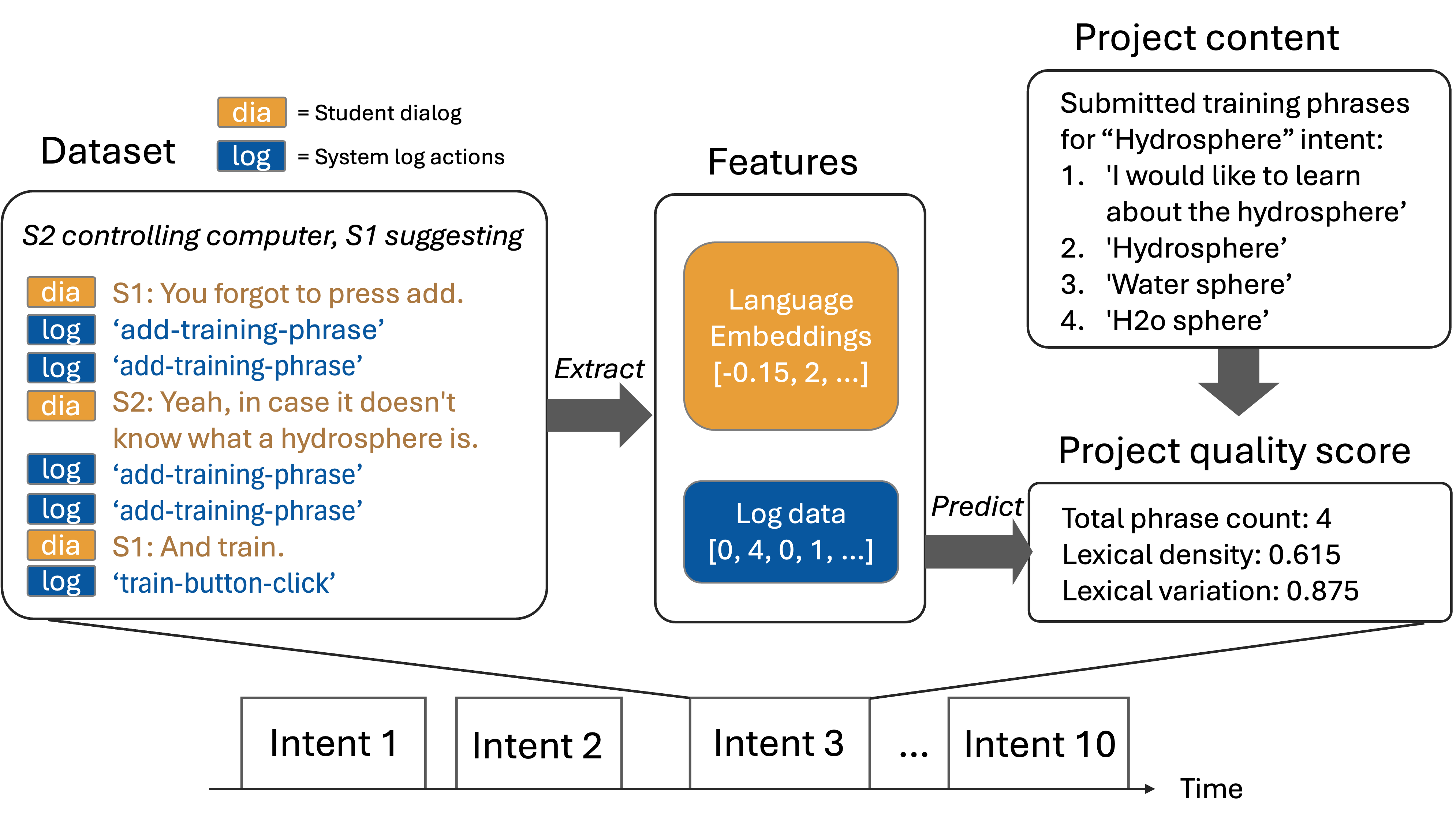}
    \caption{Overview of our multimodal predictive model, an example dataset and project outcome}
    \label{fig:model-overview}
\end{figure*}

To model relationships between system logs, student collaborative dialogue, and project outcomes, we prepared a multimodal dataset (Figure \ref{fig:model-overview}). Following Borchers et al. \cite{borchers2024using}, we synchronized dialogue transcripts with log data by aligning timestamps between the datasets, yielding an accuracy of 1 second. For each intent submission, we extracted its submitted training phrases for outcome measurement and defined \textbf{\textit{intent-working segment}}, which is the time window from when students began working on the intent until submission (based on logs). Based on an ad-hoc decision, we manually examined the segments for accuracy and removed segments that were excessively long (> 12.5 minutes, 90 percentile of the dataset) based on considerations of context length during embedding. This preprocessing step is not expected to introduce bias, as our three outcome measures were virtually uncorrelated with dialog length as measured in the total number of words ($|r|<0.05$) and total number of turns ($|r|<0.07$). The average segment duration is 193 seconds. We then extracted the synchronized student dialogues and system logs within these windows for downstream analysis (Section \ref{sec:4-analysis}).

\subsection{Outcome Measures} \label{outcome-measure}

Our unit of analysis is the \textit{intent}, a core chatbot element that handles the recognition of user queries. Students averaged seven intent training submissions per 30-minute collaborative session and 17 total submissions. We measured intent quality through three key outcomes chosen based on the learning objectives, learning curve analyses, and correlation with final project scores.

\textbf{\textit{Training Phrase Count}} measures the number of phrases input by students for training the chatbot to recognize an intent. This metric represents productivity and engagement in the iterative training process: more phrases typically indicate greater effort in improving the chatbot's performance. The relationship between chat-based activity and engagement has also been observed in past research \cite{borchers2024combining}. \textbf{\textit{Lexical Density}} captures the proportion of content words (nouns, adjectives, verbs, and adverbs) to total words in the provided phrases, which assesses linguistic richness since effective training requires substantive content over filler words \cite{kim2014predicting,lu2012relationship}. \textbf{\textit{Lexical Variation}} measures language diversity through the ratio of unique content words to total content words \cite{kim2014predicting}, important because varied phrases help chatbots recognize different user expressions. Higher variation indicates more robust training input. 

These project quality metrics align with key AI learning objectives \cite{touretzky2019envisioning,song2023ai}. For instance, students learn that simply adding similar word permutations (i.e., high training phrase quantity with little information richness, such as \verb|['hat', 'hats', 'HATS']|) is ineffective for chatbot programming.
As students iteratively generate, refine, and diversify training phrases, they engage in active problem-solving and knowledge construction \cite{park2023Imp,Lakkaraju2022}. Lexical diversity is also an important predictor of academic performance in tasks requiring writing and speaking \cite{yu2010lexical,kondal2015effects}.

\begin{figure*}[htp]
    \centering
    \includegraphics[width=0.96\textwidth]{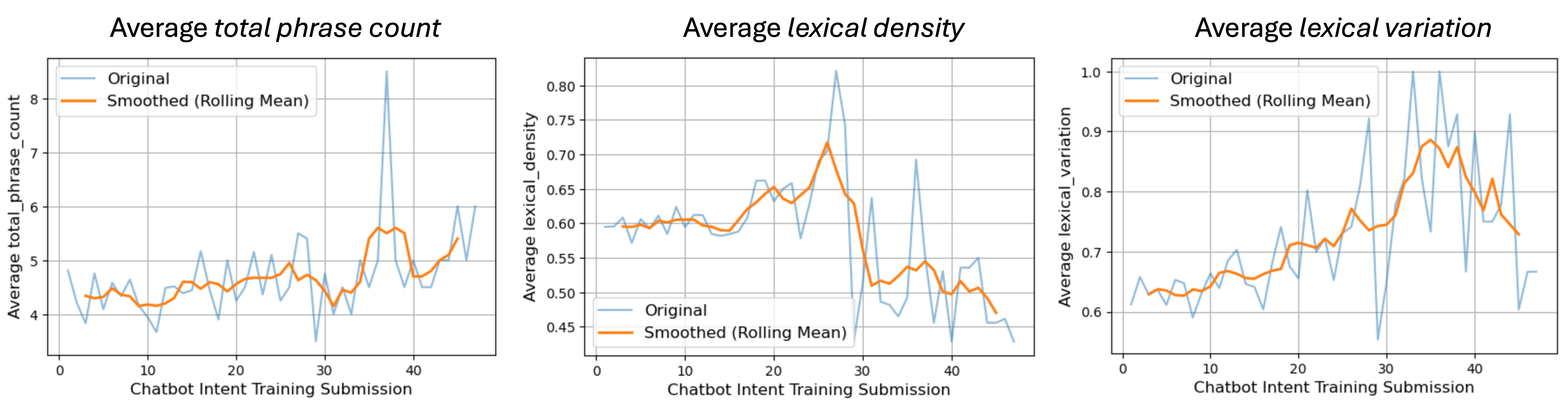}

    \caption{Learning curves for training phrase count, lexical density, and lexical variation, showing student improvements over time through smoothed running averages. Trends were less consistent for students with high submission counts (>30 intents vs. average 17), likely representing outliers.}
    \label{fig:learning_curves}
\end{figure*}

\textbf{Validation.} We validated our outcomes through learning curve analyses \cite{rivers2016learning} and correlations with three final project grades (i.e., expert-rated score aggregated across 11 dimensions using a rubric, expert-rated training phrase score, and end-user satisfaction evaluated by three independent annotators). \textbf{Learning curves} can demonstrate whether students improve on our outcome measurements as they receive indirect feedback from AMBY on their chatbot testing performance. Figure~\ref{fig:learning_curves} shows improvements across all metrics over time as students learned to add more phrases and improve lexical choices in the training data for better intent recognition (with the clearest improvement in lexical variation). For \textbf{correlations}, while none of our intent-level measures significantly \textit{correlated} with overall expert ratings--likely due to the broad scope of the overall assessment, \textit{training phrase count} significantly correlated with expert-rated training phrase scores (r(45) = 0.5, p < .001), and \textit{lexical density} correlated with end-user satisfaction score (r(45) = 0.36, p = .013), indicating richer linguistic content improves end-user experience. Our measures showed appropriate separation: moderate correlation between lexical density and variation ($r=0.28$) but minimal correlation with training phrase count ($r=0.01$ and $r=-0.03$, respectively).

\section{Analysis Methods} \label{sec:4-analysis}

\subsection{Preprocessing and Feature Engineering}

\paragraph{Log Data Features} \label{log-features}
To capture students' system interactions during chatbot development, we engineered log features reflecting engagement, timing, and strategy as inputs for our machine learning models. We computed timing-related features (six features) capturing temporal dynamics: mean, standard deviation, minimum, maximum, median, and IQR of time between actions. Event-based features (2124 features) characterized interaction patterns through event type counts and bigram/trigram sequences of the 31 unique event types (e.g., \texttt{[test-chatbot, create-new-intent, add-phrase]}). This approach of using log patterns to predict process outcomes aligns with common EDM methods where log-based learning rates relate to instructional events \cite{chi2011instructional,lin2022good}.

\paragraph{Language Features in Dialogue Transcripts}
To capture student dialogue differences, we computed numerical representations of student dialogue transcripts within a certain segment using state-of-the-art embedding models. Our early experiments showed Sentence-BERT (SBERT) embeddings \cite{reimers2019sentence} from \texttt{bert-base-uncased} performed comparably to OpenAI's \texttt{text-embedding-3-large} model \cite{wang2023improving}. Given computational efficiency and open-source considerations, we used SBERT's 768-dimensional embeddings as input features for predictive modeling, following established EDM methods \cite{zhang2024using}.

\subsection{Model Architecture, Training, and Evaluation}

We employed a feedforward neural network for regression tasks \cite{zhang2024using}, aiming to predict training phrase quality based on log data features and embeddings. The model architecture consisted of 2-4 hidden layers with ReLU activations and dropout regularization. Hyperparameters were optimized using grid search with five-fold student-level cross-validation over the following parameter space: Hidden layer configurations: \textit{[256,128], [512,256,128], [1024, 512, 256,128]} and Dropout rates: \textit{0.0, 0.025, 0.05, 0.1, 0.3, 0.5}. Model training was conducted using the Adam optimizer, and performance was evaluated using Mean Absolute Error (MAE) and AUC. Early stopping was implemented with a patience threshold of two epochs to prevent overfitting. The code is in the study's digital appendix.\footnote{\url{https://github.com/conradborchers/collaboration-edm25}}

After model training terminated, we evaluated models on a held-out test set (33\% of the data), reporting predictive performance across three feature configurations: \textbf{log only}, \textbf{dialogue only}, and \textbf{combined}. Using bootstrapped resampling, we computed 95\% AUC confidence intervals (CIs). Based on average cross-validation AUC across folds, the best model was evaluated on the held-out test set.

\section{Results}

\subsection{RQ1: Can student project quality be predicted from dialogue or log data?}
\label{sec:results_rq1}

We evaluated models trained on \textit{dialogue data} and on \textit{log data} to predict training phrase quality. Table~\ref{tab:holdout_auc} summarizes the holdout AUC. \textit{Training phrase count} was better predicted by log-based features (AUC = 0.8053) than dialogue-based features (AUC = 0.5971), suggesting system interaction logs (e.g., frequency and timing of edits) are stronger indicators of phrase production. Conversely, \textit{lexical density} was better captured by the dialogue-only model (AUC = 0.6551 vs. 0.5112), indicating student talk features capture the richness of their written training phrases. For \textit{lexical variation}, both models performed modestly (log: AUC = 0.6016, dialogue: AUC = 0.5260), with log features showing a slight advantage.

\subsection{RQ2: Does multimodal fusion improve predictive accuracy?}
\label{sec:results_rq2}

To answer RQ2, we performed early fusion, concatenating log, and dialogue features before passing them to the model \cite{sharma2020multimodal}. The results are in Table~\ref{tab:holdout_auc} (``combined'' rows). Comparing the best single-modality results, for \textit{training phrase count}, fusion improved AUC from 0.8053 (log-only) to 0.8301.\footnote{As one attentive reviewer pointed out, counting 'add-training-phrase' logs might straightforwardly reveal the number of training phrases. We performed an ablation study, recomputing these accuracies without features dependent on 'add-training-phrase', 'delete-phrase', and 'add-response'. For completeness outcome, AUCs changed to 0.6865 (log only) and 0.6759 in the combined modalities, respectively.} However, for \textit{lexical density} of training phrases, the combined model (AUC = 0.5700) did \textit{not} outperform the dialogue-only model (AUC = 0.6551), suggesting that log-based features provide limited insight into the content richness of learner input. For \textit{lexical variation} of training phrases, fusion yielded a slight AUC increase (0.6089 vs. 0.6016 log-only), but confidence intervals suggest the gain is likely not significant. Overall, multimodal fusion yields the most substantial gain for predicting \textit{training phrase count}, slightly improves the prediction of \textit{lexical variation} of training phrases, but does not help predict \textit{lexical density} of training phrases. Hence, the utility of combining modalities is outcome-dependent.

\begin{table}[ht]
\centering
\caption{Holdout AUC and 95\% Confidence Intervals for Three Outcomes Across Training Modalities}
\begin{tabular}{p{3.5cm}ccc}
\toprule
\textbf{Modality} & \textbf{Training phrase count} & \textbf{Lexical Density} & \textbf{Lexical Variation} \\
\midrule
Log Only        & 0.8053 [0.7470, 0.8604] & 0.5112 [0.4556, 0.5655] & 0.6016 [0.5418, 0.6615] \\
Dialogue Only   & 0.5971 [0.5250, 0.6671] & 0.6551 [0.5920, 0.7168] & 0.5260 [0.4579, 0.5933] \\
Combined        & 0.8301 [0.7732, 0.8822] & 0.5700 [0.5042, 0.6352] & 0.6089 [0.5438, 0.6727] \\
\bottomrule
\end{tabular}
\label{tab:holdout_auc}
\end{table}

\section{Discussion}

\subsection{RQ1: Unimodal Performance}
Our findings reveal the differential effectiveness of dialogue and log data in predicting project quality on the process level. The superior performance of log-derived features in predicting \textit{training phrase count} aligns with past work on using clickstream data to model learning trajectories \cite{karumbaiah2023spatiotemporal}, indicating behavioral engagement metrics provide critical insights into productivity and task completion. The stronger performance of dialogue features in predicting \textit{lexical density} indicates the spoken discourse appears to translate into more content-rich chatbot training phrases. This builds on research demonstrating the value of transcribed speech in detecting self-regulated learning strategies \cite{zhang2024using} and suggests that students' verbal articulation during collaboration may influence the quality of their computational artifacts \cite{grover2013computational}. 

Overall, log-data-based features were more predictive of engagement, while dialogue-based features were more indicative of the linguistic characteristics of chatbot training phrases. Future research should explore how these models learn associations between collaborative actions and outcomes, potentially incorporating interpretability methods to examine feature importance more precisely \cite{condor2024explainable}.

\subsection{RQ2: Multimodal Performance}

Multimodal fusion demonstrated mixed results in enhancing predictive accuracy. The improved prediction of \textit{training phrase count} through combined features suggests that integrating behavioral and linguistic cues provides a more comprehensive understanding of student engagement. However, for the \textit{lexical density} of training phrases, the combined model underperformed compared to the dialogue model, suggesting that adding log features may interfere with or dilute the signal from dialogue features for this task. The cross-modal interaction has also been noted in past work \cite{Hessel2020}, emphasizing that fusion requires careful consideration of interference. The modest gains in predicting \textit{lexical variation} align with past work \cite{karumbaiah2023spatiotemporal, echeverria2024teamslides}: the effectiveness of multimodal fusion depends on the prediction task. 

The results reinforce the novelty of our study's focus on open-ended project-based learning in K-12 AI education, highlighting that predicting linguistic quality in this context presents unique challenges and opportunities compared to traditional educational data mining tasks. It is evident from our results that straightforward associations between collaborative dialog and process outcomes during learning are challenging to isolate in our context (as one might conceive that predicted log inputs may be clearly visible and explicitly mentioned in dialog, and hence straightforward to predict; from our experience, these cases rarely occur). This contributes to the ongoing discourse on the role of multimodal data for learning process prediction \cite{chango2022review, borchers2024combining,zhang2024using}.

\subsection{Implications}
The findings highlight the importance of selecting appropriate data modalities for specific outcomes, contributing to multimodal learning analytics \cite{chango2022review, sharma2020multimodal}. In K-12, AI learning involving iterative chatbot design, log data, and dialogue provides complementary insights into productivity and AI understanding.

For educators assessing collaborative learning, log features offer insights into engagement and productivity \cite{karumbaiah2023spatiotemporal}, while dialogue features reveal conceptual understanding and self-regulation \cite{borchers2024combining,zhang2024using}. However, our mixed multimodal fusion results suggest that careful feature selection and tuning are essential to maximize predictive accuracy.
These insights inform future research on collaborative AI literacy tools. With sufficient accuracy, our models could monitor engagement and identify when students need scaffolding \cite{latham2010oscar, matsuda2011learning}. Dialogue analysis could help educators understand students' AI concept mastery, enabling targeted feedback and aligning with growing interest in natural language-informed interventions \cite{zhang2024using, scarlatos2024exploring, yan2024evidence}. Future work should evaluate automated transcription feasibility in real classrooms, balancing effectiveness with privacy considerations and teacher preferences \cite{yang2024leveraging}.

\subsection{Limitations and Future Work}

First, our data collection was limited to a single middle school science class with a particular collaborative task, limiting generalizability. Second, our predictive models showed only modest performance in predicting \textit{lexical variation} of training phrases, suggesting that the features and modeling techniques used might not fully capture the complexity of this outcome. These models are not deployment-ready; their current utility might lie in offline teacher analytics to inform instruction \cite{yan2024evidence}.  
Future studies could explore additional features, such as discourse-level linguistic properties or turn-taking dynamics \cite{yoo2014can}, other data fusion techniques \cite{ma2023,cohn2024multimodal} and feature importance methods (e.g., SHAP) to identify key collaborative characteristics. Third, our analysis did not separate individual student contributions within the transcripts, potentially overlooking nuances in student interactions. However, we note that, in the present study, project-based outcomes are measured and graded as group contributions. Therefore, such analyses are out of scope for this study and may require finer-grain grading and log data where interface actions are attributable to specific students, as is common in collaborative learning systems \cite{borchers2024combining}. Finally, the outcome measurement we selected may not capture certain students' unexpected behaviors (e.g., inputting a complex but irrelevant training phrase) \cite{wixon2012wtf}. While this case is rare in our dataset, future work could explore relevance-based outcome measures. 

\section{Conclusion}
This study examines the accuracy of dialogue and log data in predicting process-level collaboration outcomes in middle school chatbot development. Log features best predict student productivity, while dialogue-based features are better suited to capturing training phrase richness. Multimodal fusion improves predictions for \textit{training phrase count} and, to a lesser extent, \textit{lexical variation}, highlighting the outcome-dependent nature of multimodal models. Our research contributes to the field of multimodal learning analytics by demonstrating the potential and limitations of integrating dialogue and log data to predict collaborative learning processes and deploying related models in real-time educational settings. We also contribute insights into the interplay between student interactions and linguistic outputs in natural language programming tasks. By leveraging both dialogue and log data, educators and researchers may gain a more holistic understanding of student learning processes, informing the development of AI-powered collaborative learning tools in K-12 settings.

\section*{Acknowledgments}
This research was supported by the National Science Foundation through grant DRL-2048480. We thank Christine Fry Wise for her copy-editing of the manuscript. 

\bibliography{main}

\end{document}